Does Dark Matter Consist of Cosmic String Remnants?

Richard E. Packard
rpackard@berkeley.edu
Physics Department, University of California, Berkeley, CA

If cosmic strings have an irreducible core, then dark matter may consist of entities that are core-sized string remnants. As such they may interact solely by gravity. These **talons** may be many orders of magnitude more massive than any known fundamental particle. Due to interacting solely via gravity it would seem almost impossible to detect talons in any laboratory search.

"In talking about the impact of ideas in one field on ideas in another field one is always apt to make a fool of oneself"  R.P. Feynman, *The meaning of it all*,

One of the biggest challenges to physics is to understand the origin of the "missing" mass in the universe. This so-called dark matter, is invoked to explain several different types of astronomical observations including the flat rotation curves within spiral galaxies, the gravitational lensing around galaxies, and the clustering of galaxies in various regions of space[1]. Searches for macroscopically large, non-luminous objects (eg. planets, brown dwarfs, neutron stars, etc.) using gravitational lensing techniques, have eliminated these (so called MACHOs) as a main component of dark matter. A remaining (and widely accepted) possibility for dark matter is that the  "missing" mass in the universe is dominated by microscopic entities that carry mass but interact very weakly with other known particles.

For several decades ever more sensitive direct laboratory experiments have sought to identify dark matter entities by exploiting their possible inelastic interaction with matter in a detector. By now, the known particles included in the standard model have been eliminated[2]. Consequently an ever increasing number of more exotic entities have been proposed including: axions, sterile neutrinos, heavy neutrinos, gravitinos, anti-symmetric dark matter, GIMPs, primordial black holes, strangelets, etc. The number of possibilities grows ever longer

In this note we take an heuristic approach to add one more entity to the growing list of candidates. We present an idea, not a theory, to help explain the origin of unseen mass in the universe that cannot be detected in direct laboratory searches. Our suggestion cannot be derived or proved from known principles, certainly not by the present author. The idea is guided by two assumptions.

1. After exhaustive direct searches, no dark matter particles have been found. Therefore *we will hypothesize that dark matter particles do not interact with matter with any force other than gravity*. There are already several "particles" on the aforementioned list that might only interact via gravity and, due to the weakness of gravity, will not be registered in any detector. If direct detection is impossible perhaps the best we can hope for is to understand the origin and nature of the dark matter.

2. If one is looking for substantial amounts of non-luminous mass it is natural to be drawn toward the microscopic entities which theoretically (except for black holes) have the largest mass density ever conceived: cosmic strings (CS). This is not a new idea. Over 20 years ago Martins and Shellard considered[3] the contribution to dark matter of loops of superconducting CS. They named such structures "vortons" and concluded that for the superconducting CS that they considered, it would be difficult to dynamically stabilize small particle–like loops. Other authors have also considered dynamical means of stabilizing vortons so they do not "evaporate" in the earliest cosmic moments[4]. Our suggestion here is that *if CS have an irreducible core, the end point of string evolution will necessarily consist of core size bits of twisted space-time, distributed in the cosmos to provide the missing dark matter.*

The stability of this entity comes from the core structure, not from dynamics. An irreducible core is not included in the several models of CS that have been treated exhaustively in the theoretical literature. We cannot prove such a model and do so only to force the existence of the entities we describe herein. This will not be the first time that physics requires a model outside of the accepted paradigm in order to explain observational data[19].

Although there has been exhaustive theoretical work dealing with CS it is important to recall that CS are presumed to exist in an energy epoch many orders of magnitude greater than where there is any existing empirical evidence. Therefore the models for the properties of CS, including those so thoroughly treated previously, must be taken as speculative.

This note will first review the general idea of a cosmic string. Then, because CS are somewhat analogous to vortex lines in superfluid helium[5], we will explain how superfluid vortices evolve in time into irreducible entities. It is this analog that is the basis for the talon concept described below. We then will discuss some relevant orders of magnitude that indicate the number density of talons required to satisfy dark matter observations.

The idea of cosmic strings (CS) arose from the seminal concept of T. Kibble[6] in 1976. In the standard big bang model, the rapidly cooling universe passes though a series of symmetry breaking transitions wherein the known (non gravitational) forces of nature (the strong interaction, the weak force of beta decay, and electromagnetism) are initially unified at energies above about $10^{15}$GeV, the era treated by grand unification theory (GUT). As the universe cools the interactions separate into their present distinct forms. These transitions have been described as being mathematically similar to some second order phase transitions in condensed matter, such as the normal to superfluid transition in liquid helium and the normal to superconducting transition in some metals. When cooled rapidly though the transition temperature, these condensed matter transitions characteristically exhibit topological defects in the lower temperature state: quantum vortices in superfluids and magnetic flux lines in superconductors. By analogy Kibble suggested that space-time itself might contain similar topological defects, the most likely being linear structures referred to as cosmic strings[7,8].

Most attention related to cosmic strings has been focused on the transition when the strong force is believed to have separated from the electroweak interaction. It is believed that this occurred at a high critical temperature (energy) of $E_c \sim 10^{15}$ Gev. This energy sets the scale for remnant mass and size. The transition is believed to have occurred near the end of the epoch of the putative cosmic inflation. If the CS were created before the inflation period ended, their characteristic energy might be appreciably less but still at energies much greater than observational physics has probed.

Searches for evidence of extended cosmic strings have involved several observational techniques including pulsar timing, analysis of the cosmic microwave background (CMB) and microlensing. As yet there is no observational evidence[9] for extended cosmic strings. Therefor if one accepts the premise that spatially extended CS existed in the earliest moments of the universe, one might conclude that these primordial lineally extended defects in space-time, must have vanished before the earliest observable epoch. Here we suggest how compact relics of CS might have survived and may be a central component of dark matter?

The basis of this idea is determined by the structure of the CS "core". In accepted models this core is not topologically stable. This permits a CS to ultimately evaporate as it breaks into ever smaller loops[5]. Theorists pursued models wherein a small loop of CS would be dynamically stabilized by, for example, supercurrents in the loop. Such dynamically stabilized loops have been named vortons. In most reports such stable "vortons" have not been found. In the following we point out that if the core is irreducible, as in the case of superfluid vortices, then core-sized loops can be stabilized structurally, rather than dynamically.

To understand the possible role of CS related to dark matter it is important to first understand some features of superfluid helium vortices[10]. Superfluid helium is described by Landau's two fluid model[11] wherein the fluid is composed of two interpenetrating components: a superfluid fraction which is a macroscopic quantum state and a normal component that consist of the thermal excitations of the liquid. The normal fluid fraction decreases rapidly with falling temperature, from 100% at the superfluid transition, $T_c$, to essentially zero when $T/T_c$ is less than $\sim 0.5$. Correspondingly the superfluid fraction which is zero at the transition is essentially unity below this temperature.

Quantum vortices are observed to occur in the superfluid fraction. The quantized vortex line structure is described having an azimuthal superfluid velocity falling off inversely with distance from the center. To avoid a velocity singularity at the origin, an atomic-size core is assumed to be normal, i.e. not superfluid, irrespective of the temperature. These vortices have a superfluid circulation equal to Planck's constant divided by the atomic mass of helium. Many types of experiments confirm this model. For the discussion herein the important aspect of the normal core is that it is irreducible. **No vortex element smaller than the core size can exist. Vortices cannot simply evaporate.**

Superfluid vortices have been well studied in the laboratory for over half a century. Numerical simulations of their motion have been directly confirmed by experiments[12]. Figure 1 is an example[13] of a simulation showing the vortex line distribution in the limit of low temperature when there is no normal fluid present (outside the core) to damp the motion. The vortex lines develop wiggles and twists. A key feature is that when two line segments closely overlap there is a reconnection event that produces a loop separated from the remaining line. The macroscopic loop itself continues to develop wiggles and more twists, producing smaller loops whenever a reconnection occurs. This cascade (similar to a Richardson cascade in classical turbulence[14]) continues downward to ever-smaller scales. Simulations cannot follow the motion down to loops of atomic dimensions, and the structure of the atomic-size normal core is just a simple model. It is assumed that eventually the original energy in the extended vortices is dispersed in two channels: 1. Terminal loops comparable to the core size typically $\sim 10^{-10}$m. and 2. Sound radiation[15] triggered during the rapid relaxation of the sharp cusp associated with a reconnection event. The terminal loops, which are the remnants of the original vortex line tangle, are assumed to eventually collide with the walls bounding the superfluid and become part of the thermal excitations. Indeed the higher temperature superfluid excitations, named rotons, may be the smallest vortex rings.[16].

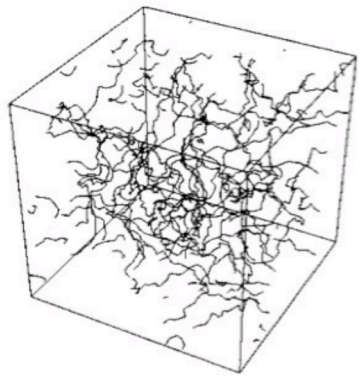

Figure 1. A simulation[13] of a superfluid vortex tangle in the absence of normal fluid, i.e. at low temperatures. In the simulation, loops that break off after reconnections can vanish by passing through the simulated bounding walls.

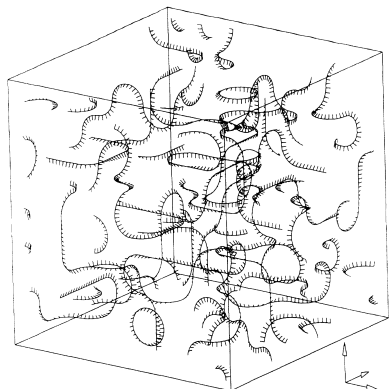

Figure 2. A simulation of cosmic strings evolution. See reference 17 for details.

Figure 2 is a simulation of a CS tangle. The similarity between the vortices and the CS is striking. Several impressive simulations[17] describing the evolution of cosmic strings have displayed dynamics similar to the dynamics of superfluid vortices.  In particular the strings "wiggle", and sections "interconnect" with neighboring sections when two string segments cross. These reconnections result in loops forming, which, themselves evolve through wiggles and further reconnections to smaller and smaller loops. The loops cascade downward in size. Simulations cannot follow this cascade to ever decreasing size, both because of the limitations of computation time, but also due to a lack of knowledge of the structure of the core of the string. It is this similarity to superfluid vortices that is guiding the model suggested herein.

One can ask: For CS is there ultimately a smallest loop or structure, characteristic of the core dimensions? Then further evolution would cease and we are left with core size "loops" of string that are metastable on the time scale of the age of the universe[18]. Small superfluid rings will vanish when they collide with a wall. By contrast, for endpoint cosmic string loops, there are no walls to collide with, and these small entities might drift forever through the universe. We call the end point remnants of primordial cosmic stings, "talons". For clarity, we differentiate this term from a "vorton" which is a loop of CS stabilized by dynamical interactions rather than the talons which are stabilized structurally.

The properties of a talon might be determined by the energy scale of the initial phase transition.  If CS appear after the inflation period, at the end of the GUT era, $\sim 10^{15}$GeV , a talon mass might  be 15 orders of magnitude greater than a proton.

The talon size would be set by the core dimension, which in turn is determined by the inverse of the energy of the phase transition.  Thus a talon might be 15 orders of magnitude smaller than a proton.  The density is only a couple of orders of magnitude below that of a black hole.

A talon will only interact with ordinary matter via gravity. Therefore searches relying on other particle interactions will not detect talons. This feature, which is characteristic to any purely gravitationally interacting particle. Is consistent with the null results of laboratory direct dark matter searches and is one driving motivation for this talon hypothesis.

The talon's enormous mass, possibly 15 orders of magnitude greater than baryons, implies that a number density  correspondingly 15 orders less than baryonic density yet would play a gravitational role comparable to the visible mass in the universe. Thus, not only are talons undetectable thru inelastic interactions but in addition, their flux thru any region would be very small. The dark matter mass density within galaxies is estimated to be $\sim 0.3$GeV/cm$^3$. Dark matter particles are estimated to have a velocity $\sim 200$km/s. If dark matter consists of talons with mass $\sim 10^{15}$GeV, their number density would be $\sim 10^{-15}$/cm$^3$. A typical number flux would be $\sim 2\times 10^{-8}$/cm$^2$s. Thus not only would talons be undetectable because gravity is such a weak force but in addition, the probability of a talon passing thru a human-scale detector would be very small. Only on the galactic scale would the presence of talons be noticeable.

**Discussion**

Here we have made the ad hoc hypothesis that CS are similar to quantum vortices in that the core is irreducible. This is an ad hoc and unprovable assumption made to guarantee that CS will eventually degrade to the end point structures we call talons. There are several several cases where a new assumption was required to explain an observational mystery[19]. One does not make such assumptions casually but the pervasiveness of a mystery forces physicists to occasionally try new approaches, however radical.

It is important to mention that to have confidence in models about nature on a length scale of $10^{-30}$m is really an act of faith. At present we know nothing observationally about physics on the length scale of a talon. To extend ideas originating from an established regime into a regime observationally inaccessible by so many orders of magnitude, often shows those ideas require modification.

At the time of the GUT transitions, CS are believed to be moving at relativistic speeds comparable to the speed of light[8]. If talons are the end point entities of those strings one must ask how they decelerated to speeds like a few hundred km/sec, as required to form the dark matter halos that dominate in spiral galaxies. A possible energy loss mechanism would be the gravitational analog of elastic coulomb scattering. A talon can scatter off another particle through the attractive gravitational interaction. The most effective energy transfer occurs for scattering off particles with mass comparable to that of the talon. In such scattering event some small amount of energy will also be lost via gravitational bremsstrahlung radiation.

Eventually talons may be trapped in the potential well of clumps of talons and baryonic matter, thus providing the dark matter that characterizes the flat galactic rotation curves.. If talons orbit at speeds comparable to the speed of stars in the arms of galaxies they will move in concert with the neighboring matter and scattering and energy loss will diminish.

Given the weakness of the gravitational interaction it would seem unlikely that a talon scattering event would register in any detector. Thermal noise, even at millikelvin temperatures, would likely swamp the recoil energy of a target particle. Further, the constrictive impulse produced by a talon passing through a resonant crystal could only excite the resonant modes of that crystal if the speed of the talon is comparable to the speed of sound in the material. But even in a hard material such as diamond, the speed of sound is two orders of magnitude below the speed of dark matter (i.e. a few hundred km/s) in the halos of galaxies.

The purpose of this note is to add one more entity to the long list of dark matter candidates. The heuristic idea presented herein is that if cosmic strings existed and if those strings have an irreducible core, than their evolution would terminate with remnant, core sized entities that might form dark matter.

We may never know if any suggestions, made thus far or in the future, form a realistic explanation of the observations that support the dark matter idea. If talons, or some undetectable entity like them, dominate the cosmos, physics will be in crisis. To claim that the universe is dominated by things we can never directly detect is to replace fact with believe. That is not the physics that exists heretofore. .

**Acknowledgement:** I thank S . Ominsky for his encouragement. Early aspects of this work was supported in part by the NSF Division of Material Science.

---

[1] An alternative explanation to these observations may lie in an extended understanding of gravity. For this note we will assume that dark matter does exist and suggest a possible model for its origin. A modification of gravity is a much more difficult task than to pursue the simpler assumption that general relativity is complete and dark matter exists

[2] For a recent review of WIMP searches see, Roszkowski, L., arXiv:1707.06277v2[hep-ph]; also see D. Bauer, Searching for Dark Matter, American Scientist, **106**, 303 (2018).

[3] C. J. A. P. Martins and E. P. S. Shellard, Vortons: Dark Matter From Cosmic Strings, *The Non-Sleeping Universe* pp 325-326, Conference proceedings, Springer (1997); see also by the same authors, Vorton Formation, Phys. Rev D, **57**, 7155 ( 1998)

[4] J. Garaud, E.Radu, M. Volkov, Stable Cosmic Strings, Phys. Rev. Lett., **111**, 171602 (2013)

[5] R. L. Davis and E.P. S. Shellard, Global Strings and Superfluid Vortices, Phys. Rev Letters, **63**, 2021 (1989). These authors assume that the magnitude of the Higgs field vanishes at the string center. This allows a small loop to evaporate. Here we suggest that the core may be irreducible and a small ring will not evaporate.

[6] Kibble T W B *J. Phys.* A: *Math. Gen.* 9re 1387 (1976)

[7] For a review of cosmic strings see Copeland, E.T. and Kibble, T.W. B., Cosmic Strings and Superstrings, Proc. Roy. Soc A, doi:10.1098/rspa.2009.0591

[8] An earlier and more complete review is found in A. Vilenkin and E.P.S. Shellard, Cosmic Strings and Other Topological Defects, Cambridge University Press. 1994

[9] E. Jeong, et al, Probing Cosmic Strings With Satellite CMB Measurements, Journal od Cosmology and astroparticle physics, 09(2010) 018

[10] *Quantized vortices in HeII*, R.J. Donnelly, Cambridge University Press

[11] D.R. Tilley and J. Tilley, Superfluidity and Superconductivity, third edition, , Institute of Physics Publishing, Bristol and Philadelphia (1990)

[12] See for example, K.W. Schwarz, "Unwinding of a single quantized vortex from a wire", Phys. Rev. B., **47**, 12030 (1993). Also, S. Fujiyama et al, Phys. Rev. B, **81** 18051, (2010)

[13] , M. Tsubota, A., Tsunchiko, and M. Nemirovskii, Dynamics of vortex tangle without mutual friction in HeII, Phys. Rev. Lett., **62**, 11751(2000)

[14] L. F. Richardson composed the limerick: "Big whirls have little whirls that feed on their velocity, and little whirls have lesser whirls and so on to viscosity."

[15] Sound emission due to superfluid reconnections, M. Leadbeater, et al, Phys. Rev. Lett., **86**, 1410 (2001)

[16] The statement that "a roton is the ghost of a vanishing vortex ring." is attributed to L. Onsager by R.J. Donnelly in ˆ*Quantum Statistical Mechanics in the Natural*

*Sciences*, Vol. 4 of *Studies of the Natural Sciences,* Eds. S.L. Mintz and S.M. Widmayer, pg. 359, (1974)

[17] B. Allen and E.P.S. Shellard, Cosmic String Evolution:A Numerical Simulation, Phys. Rev. Letter, **64**, 122 (1990)

[18] It might be useful for the reader to consider the case of an ordinary piece of string with a loose overhand knot placed along it. If the ends of the string are pulled tight the knot decreases in size until it jams when the knot's diameter is comparable to the string's diameter. There can be no smaller knot.

[19] Those instances included quantizing energy to explain black body radiation, considering light to be made of particles to explain the photoelectric effect, giving the electron quantized spin to explain atomic fine structure, quantizing angular momentum to explain the spectra of atomic hydrogen and invoking dark matter itself to explain intergalactic stellar motion.